\documentstyle[12pt,moriond,psfig]{article}
\newcommand{\Ha}{H$\alpha$}
\newcommand{\Hb}{\ifmmode {\rm H}\beta \else H$\beta$ \fi}

\newcommand{\mup}{$M_{\rm{up}}$}
\newcommand{\Zw}{I~Zw~18}
\newcommand{\hii}{H~{\sc ii}}
\newcommand{\Heii}{He~{\sc ii} $\lambda$4686}
\newcommand{\Oiii}{[O~{\sc iii}] $\lambda$5007}
\newcommand{\oiii}{[O~{\sc iii}]}

\newcommand{\msun}{\ifmmode M_{\odot} \else M$_{\odot}$\fi}
\newcommand{\rsun}{\ifmmode R_{\odot} \else R$_{\odot}$\fi}
\newcommand{\lsun}{\ifmmode L_{\odot} \else L$_{\odot}$\fi}
\newcommand{\zsun}{\ifmmode Z_{\odot} \else Z$_{\odot}$\fi}
\newcommand{\heii}{He~{\sc ii}}

\begin{document}
\heading{MASSIVE STAR POPULATIONS IN I ZW 18: A PROBE OF STELLAR EVOLUTION
IN THE EARLY UNIVERSE}

\author{D. Schaerer $^{1,2}$, D. de Mello $^{2}$, C. Leitherer $^{2}$, J .Heldmann
 $^{3}$} {$^{1}$ Observatoire Midi-Pyr\'en\'ees, F-31400 Toulouse, France.}
 {$^{2}$ STScI, Baltimore, MD 21218, USA.
 $^{3}$ Colgate University, Hamilton, NY 13346, USA.}

\begin{moriondabstract}
We present a study of the gaseous and stellar emission in I Zw18,
the most metal-poor star-forming galaxy known.
Archival HST WFPC2 and FOS data have been used to analyze the spatial 
distribution of  \oiii, \Ha, and \Heii. The latter is used
to identify Wolf-Rayet stars % recently found by ground-based spectroscopy
and to locate nebular
\heii\ emission. Most of the \heii\ emission is associated with the
NW stellar cluster, displaced from the surrounding shell-like
\oiii\ and \Ha\ emission.
We found evidence for \heii\ sources compatible with 5--9 WNL stars
and/or compact nebular \heii\ emission as well as residual diffuse
emission.
New evolutionary tracks and synthesis models at the appropriate metallicity
predict a mass limit $M_{\rm WR} \approx$ 90 M$_{\odot}$ for WR stars to become 
WN and WC/WO.
The observed equivalent widths of the WR lines are in good agreement with 
an instantaneous burst model with a Salpeter IMF extending
up to $M_{\rm up} \sim$ 120-150 \msun.
Our model is also able to fully reproduce the observed 
equivalent widths of nebular \heii\ emission due to the presence of WC/WO stars.
This finding together with the spatial distribution of nebular 
\heii\ further supports the hypothesis that WR stars are responsible 
for nebular \Heii\ emission in extra-galactic \hii\ regions. 

Finally we discuss the implications of our results on stellar mass loss, chemical yields,
final stellar masses, and the ionizing flux of starburst galaxies at very low 
metallicities.

\end{moriondabstract}

\section{Introduction and motivation}
The main aims of our work are the following:
\begin{itemize}
\item {\sl Probe the evolution of massive stars in low metallicity systems.}
Since environments with massive stars at metallicities $Z < 1/10 $ \zsun\ are not 
available in the Local Group we use star-forming regions and super star clusters
in BCDs. 
It is important to constrain the populations of massive stars in
different environments, since their evolution is dictated by mass loss whose
metallicity dependence is not well known.

As a consequence all predictions related to massive stars depend on the adopted 
mass loss prescriptions. E.g.\ chemical yields are $Z$ dependent (\cite{m92}), 
and the P-Cygni lines detected in several high redshift galaxies (e.g.\ \cite{st96})
depend on the mass loss properties.

To constrain the evolution we analyse the Wolf-Rayet (WR) star content since these
stars represent bare stellar cores revealed by mass loss. From the WR and O star
content we can also derive constraints on the upper mass and the slope of the
IMF (cf.\ \cite{s96}, \cite{l98}) % Leitherer, these proceedings).

\item {\sl Explain the origin of nebular HeII emission frequently observed in
low metallicity extra-galactic HII regions.}
The nature of this emission remained puzzling until recently (cf.\ \cite{g91}
and \cite{s96}, \cite{s97}) and indicates a harder ionizing spectrum than 
commonly thought.
\end{itemize}

\section{Observations of I Zw 18}
Signatures of WR stars were detected very recently in I Zw 18 by\cite{i97} and
\cite{l97}.
Although their spectra differ considerably they clearly establish the presence 
of WR stars in the most metal-poor galaxy known. Their discovery raises, however,
a few interesting questions, namely:
1) Can the observed WR population be reproduced by stellar evolution models
at the metallicity of I Zw 18, and if so what mass loss rates are required ?
2) Is the observed nebular \heii\ $\lambda$4686 emission due to WC stars as suggested
by \cite{s96} ?
To answer these questions we have analysed HST images of I Zw 18 and calculated
new stellar evolution tracks and synthesis models. A detailed account of this
work is given in \cite{d98}

\begin{figure}[htb]
%\centerline{\psfig{figure=../../wrgalaxies/izw18/figs/fig2.eps,height=8cm}
\centerline{\psfig{figure=fig2.eps,height=8cm}
\psfig{figure=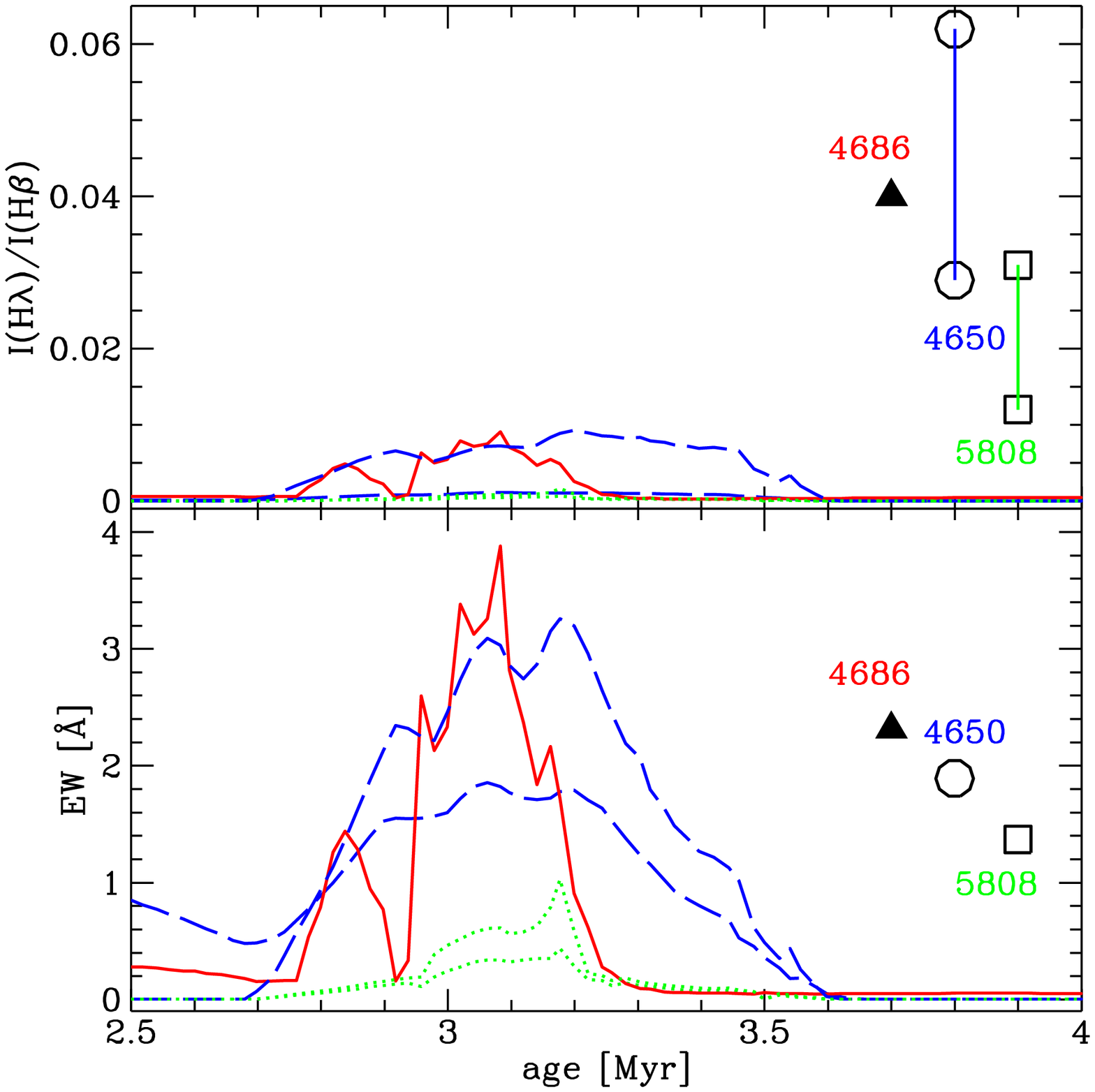,height=8cm}}
\caption{{\protect\small {\em Left:} WFPC2 V (F555W) image of \Zw. 
The rectangle delineates the area in the NW region where
the continuum-free helium sources were detected. The darkest pixels in the 
helium
map are above the 3$\sigma$ level. WR, WR?, and WR?? identify helium 
sources with fluxes equivalent to 3, 0.7, and 0.4-2. WNL stars, respectively.  
{\em Upper right:} Comparison of predicted relative line intensities
of the WR bumps (4650 \AA: dashed, 5808 \AA: 
dotted) and nebular \Heii\ (solid) with observations of \cite{l97} and \cite{i97}.
Instantaneous burst model with Salpeter IMF.
{\em Lower right:} Same as upper panel for equivalent widths.
% compared to the data from \cite{l97}.
Note the good agreement between model and observations.}}
\label{fig_heii}
\end{figure}

We used archival HST WFPC2 imaging and FOS spectroscopy (proposal ID 5309, 5434, 6536)
to construct continuum free images of \Ha, \Oiii, and \Heii.
Since the expected flux levels in \Heii\ are close to the background and the noise
level of the detector,  careful cosmic ray and hot pixel removal, and back ground 
substraction were done in addition to the standard pipeline processing (see \cite{d98}).
Several procedures were experimented to construct the continuum free line maps.
From this we conclude that our \Heii\ map (see Fig.\ \ref{fig_heii}) is
well-suited  to detect any excess \heii\ emission. As a conservative error we
adopt 30 \% for its absolute flux calibration.

The \heii\ map shown in Fig.\ \ref{fig_heii} reveals several compact \heii\
sources centered on the NW cluster of I Zw 18. The significance of the 
individual pixel detections is as follows: 62 pixels with $>$ 2 $\sigma$,
25 with $>$ 3 $\sigma$, and 12 with $>$ 3 $\sigma$. 
Residual diffuse emission over a 9.9 arcsec$^2$ region is also found.

From the line flux the most significant sources are either compatible with
$\sim$ 0.5 -- 3 WNL stars (using the line luminosity from \cite{sv98}) or
with very compact nebular sources (similar to the LMC nebula N44C
with an angular diameter $\sim$ 0.03$''$; cf.\ \cite{g91}).
The spatial analysis of the \heii\ sources, possible thanks to the high 
resolution of HST, reveals the following:
Whereas the \heii\ is centered on the stellar cluster the maximum of \Ha\ and 
\Oiii\ emission is clearly displaced from the cluster in a shell like
structure (see \cite{d98}).
The spatial distribution of \Heii\ is also compatible with the ground-based
data from \cite{l97} and \cite{it98} who find a {\it spatial correlation}
between nebular \heii\ and the WR features, suggestive of a link between
the WR stars and nebular \Heii\ emission.

\section{Comparison with synthesis models}
For a quantitative comparison between observations and stellar evolution models
we have calculated new evolutionary tracks at $Z=1/50$ \zsun\ adopting either
the high mass loss rates as in \cite{m94}, or $\dot{M}$ from the wind momentum 
- luminosity relation (\cite{lc96}).
Interestingly, for the purpose of the present work, both prescriptions yield
(due to the ``saturation effect'' at high luminosities)
the same results: 1) the WR mass limit at $Z=0.0004$ is $M_{\rm WR} \sim$ 
90 \msun, and
2) the models predict both WN and WC/WO stars.

We have used these tracks in the synthesis models of \cite{sv98} to predict
the strength of the WR and nebular emission lines. The results are compared
in Fig.\ \ref{fig_heii} (right) to the observations of \cite{i97} and \cite{l97}. 
As shown in the upper panel the observed intensity ratios of the WR 
and nebular \Heii\ lines are considerably larger than the predictions.
This can easily be explained by the spatial offset between these emission 
features and the \Ha\ emission (see \cite{d98}).
On the other hand the predicted equivalent widths (lower panel) of the WR lines
are in good agreement with the observations.
The observed WR population in I Zw 18 can thus well be reproduced with a
Salpeter IMF and an upper mass cut-off \mup\ $> M_{\rm WR}$ of typically 
120-150 \msun.
At the same time the observed nebular \Heii\ emission (W(4686) $\sim$ 2.4 \AA)
is naturally explained by the presence of WC/WO stars as suggested by \cite{s96}
and \cite{s97}.

\section{Implications}
Combining the results of various studies on WR and O stars populations
in the Local Group and in young starbursts several conclusions can be drawn.
First massive star evolution models with high mass loss rates (cf.\ \cite{m94}),
and/or to a certain extent also additional mixing (\cite{mm97}),
are clearly favored from a variety of studies including: the properties of individual
WR stars (\cite{mm94}), analysis of WR and O star populations in Local Group
Galaxies (\cite{mm94}) and in WR galaxies with $-0.9 \le {\rm [O/H]} \le 0.2$
(\cite{mc94},\cite{s96},\cite{setal96},\cite{c98}). The present study now extends this
result to the galaxy with the lowest metallicity known to date
(I Zw 18: [O/H] $\sim -1.7$).

The effect of metallicity dependent mass loss on pre-SN core masses
and stellar yields has been discussed by \cite{m92}.
If taken at face value the above conclusion implies that compared
to \cite{m92} the chemical yields of He are even larger and those of 
CO smaller at low metallicities. 
In particular observations of the relative WN and WC star populations
in WR galaxies
which have only recently become available (see \cite{setal96}, \cite{c98})
should allow to place constraints on the relative yields of He and CO
from massive stars.
We also note that the favored high mass loss models lead to
final (pre-SN) masses which are generally smaller than the results
obtained by \cite{p98}.

Another implication from our study concerns the far UV spectrum
of young galaxies. As can be seen from the sample of low metallicity
galaxies (mostly BCD) of Izotov and collaborators, approximately 
$\sim$ 70 \% of their objects show nebular \Heii\ emission.
This indicates that the majority of low [O/H] emission line
objects have a fairly hard far UV spectrum, which we attribute to the
presence of early type stars (cf.\ above).
This finding may prompt a re-examination of the importance of
young galaxies to the ionization of QSO absorption line
systems and the ionization of the integalactic medium (cf.\
\cite{g91}, \cite{d98}).

As illustrated in this work studies on massive star populations in
metal poor galaxies and the origin of nebular \heii\ emission
provide interesting constraints on stellar evolution in the
early universe.

\noindent
{\small {\sl Acknowledgments}
DS is grateful to the Swiss NSF and the ``GdR galaxies'' for 
financial support. 
}

\vfill
\end{document}